\documentclass[slac_one]{revtex4}
\usepackage{graphicx}
\usepackage{fancyhdr}
\begin{document}

\title{{\small{Hadron Collider Physics Symposium (HCP2008),
Galena, Illinois, USA}}\\ 
\vspace{12pt}
Overview of the theory of $\mathbf W$/$\mathbf Z$ + jets and heavy flavor } 

%

\author{J. M. Campbell}
\affiliation{University of Glasgow, Glasgow, G12 8QQ, UK}

\begin{abstract}
I review the status of theoretical predictions for events
containing a $W$ or $Z$ boson and jets,
one or more of which may include heavy quarks.
Special attention is paid to comparisons between different theoretical
approaches and with the latest experimental data.
\end{abstract}

\maketitle

\thispagestyle{fancy}


\section{INTRODUCTION} 

$W$ and $Z$ bosons are currently observed at a
very high rate at the Tevatron, with cross sections of the order of
a nanobarn, and will be even more copiously produced at the LHC.
In such hadronic environments, the  vector bosons are observed in
association with strongly-interacting particles, which are 
understood in terms of the radiation of additional gluons and quarks
in Quantum Chromodynamics (QCD). Much of this radiation is {\em soft},
corresponding to transverse energy deposits of a few GeV or less.
However, producing a significant deposit of energy that may be
interpreted as a jet of particles originating from a single hard gluon
is still fairly common.

\begin{figure}
\includegraphics[width=8.1cm]{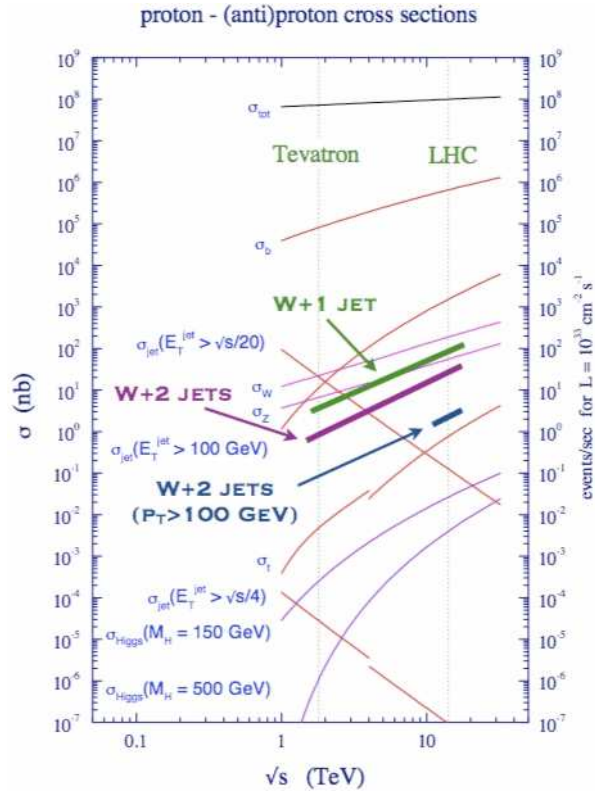}
\caption{\label{wjt:xsecs}
Cross sections for a range of basic processes as a function of the
center of mass energy, $\sqrt s$
(adapted from Ref.~\cite{Campbell:2006wx}). The lines indicated
for $W+1$,$2$ jets correspond to jets with $p_T>20$~GeV and
$|\eta|<2.5$.}
\end{figure}

Na\"ive power counting, with additional gluon radiation suppressed
by a factor of $\alpha_S(M_V) \approx 0.1$, is borne out by more
detailed calculations, as illustrated in Figure~\ref{wjt:xsecs}.
The cross section for
producing a $W$ boson together with jets of $p_T>20$~GeV drops by an
order of magnitude as each additional jet is considered. The event
rate for producing a $W$ and two such jets is still three orders of 
magnitude larger than the expected total cross section for
production of a
Higgs boson of mass $150$~GeV. At the LHC, events with two very hard
jets, of transverse momenta larger than $100$~GeV, will still be
relatively commonplace. With a cross section of a few nanobarns,
they will be produced as frequently as the much softer $W+2$~jet
events currently are at the Tevatron.

When folding in the decays of the $W$ and $Z$ bosons, one sees
that these types of events can lead to substantial backgrounds to
many searches at the Tevatron and the LHC. The leptonic decays
of the $W$ and $Z$ lead to signatures consisting of missing energy, 
lepton(s) and jets. The presence of a lepton is especially
important since it can be used to tag the event 
and thus it is in such decay modes that these channels are most
important. The recent evidence for single top production at the 
Tevatron~\cite{CDFsingletop,Abazov:2008kt} has relied crucially
on a good understanding of the $W+2$~jets background
when one or both of the jets may contain bottom quarks. Furthermore,
supersymmetry and other models for new
physics provide a plethora of signals identified via missing energy and
high transverse momentum jets. To pick just one example, a relatively
light Higgs boson produced in the vector boson fusion (VBF) channel
may be identified via its decay to tau leptons. The observed final
state, $\tau^+ \tau^- +$~jets, is easily mimicked by $Z$ boson
production~\cite{Rainwater:1998kj}.

In order to have good control of these backgrounds across a wide
range of detector acceptances, it is crucial that they are both well
understood and accurately simulated theoretically. This goal can
only be achieved by systematic advances on two fronts: improvement of
the theoretical calculations and validation of calculational
tools against existing data. As well as the immediate rewards that may
be reaped from such a programme, in the longer term these studies can
serve as benchmarks for similar future analyses. In particular,
plentiful production of top quark pairs at the LHC will likely
necessitate systematic studies of $t{\bar t} +$~jet events along
similar lines.

\section{$\mathbf W$/$\mathbf Z$ + JETS}

Until recently, the theoretical tools available for comparison with
hadron collider data have fallen into two camps: fixed order
parton-level calculations and parton shower approaches based on the
factorization properties of QCD amplitudes.

Working in fixed order
in perturbation theory is by now straightforward at leading order (LO),
but already limited at next-to-leading order (NLO)~\cite{FMaltonitalk}.
At next-to-next-to-leading order (NNLO) very few calculations have
been performed, with existing ones limited to final states 
containing only one particle. In addition, the capability of such
calculations to model the final states found in the experiments is
very limited. At LO each parton in the calculation corresponds to a
single jet, with one additional parton per jet allowed for each
successive order. Moreover, the number of final state particles that can
currently be included in a full NLO calculation is three. Despite these
flaws, the great advantage of these calculations is that they provide
a reliable normalization of the cross section (NLO and beyond) whose
uncertainties can be sensibly estimated (starting at NNLO).

On the other hand the parton shower approach is ideally suited to
simulating final states similar to those found in the data which
allows, for example, a full analysis of detector effects. This is
possible thanks to the additional radiation that is generated
stochastically in the shower, after starting from a given (usually 1 or
2 particle) hard final state. The theory underlying such an approach
is based on the factorization of the perturbative matrix elements in 
the limits of soft and collinear partons. Although the effects of such
soft physics are taken into account and well-modelled, hard or large
angle radiation is in general poorly described. In addition,
this approach is generally implemented only using LO matrix elements,
limiting the trustworthiness of the predictions accordingly.

Recently there has been progress on both fronts. At fixed order there
have been many developments aimed at automating
NLO calculations so that they may be applied to higher multiplicity
final states~\cite{FMaltonitalk,Bern:2008ef}. Regarding parton showers,
there have been two major developments in recent years. The first,
generically known as {\em matching}, supplements the traditional parton
shower by including hard matrix elements with more partons. In this way, the
poor approximations usually employed by a parton shower in hard
configurations are eliminated by using the exact matrix elements.
The original implementations of such
schemes~\cite{Catani:2001cc,Mangano:2002ea} have recently spawned
further versions~\cite{Schwartz:2007ib,Bauer:2008qh,Bauer:2008qj},
although the basic
principle remains the same. The arena of $W/Z$+jet simulation has been
an ideal testing ground for the idea of matching in parton showers, as
we shall discuss shortly. Finally, a major advance in this area has been
the inclusion of NLO matrix elements in a parton
shower~\cite{Frixione:2003ei}.
At present this implementation does not address the issue of $W/Z$+jet
production, although it is not infeasible in the near future.

\begin{figure}
\includegraphics[width=5.3cm]{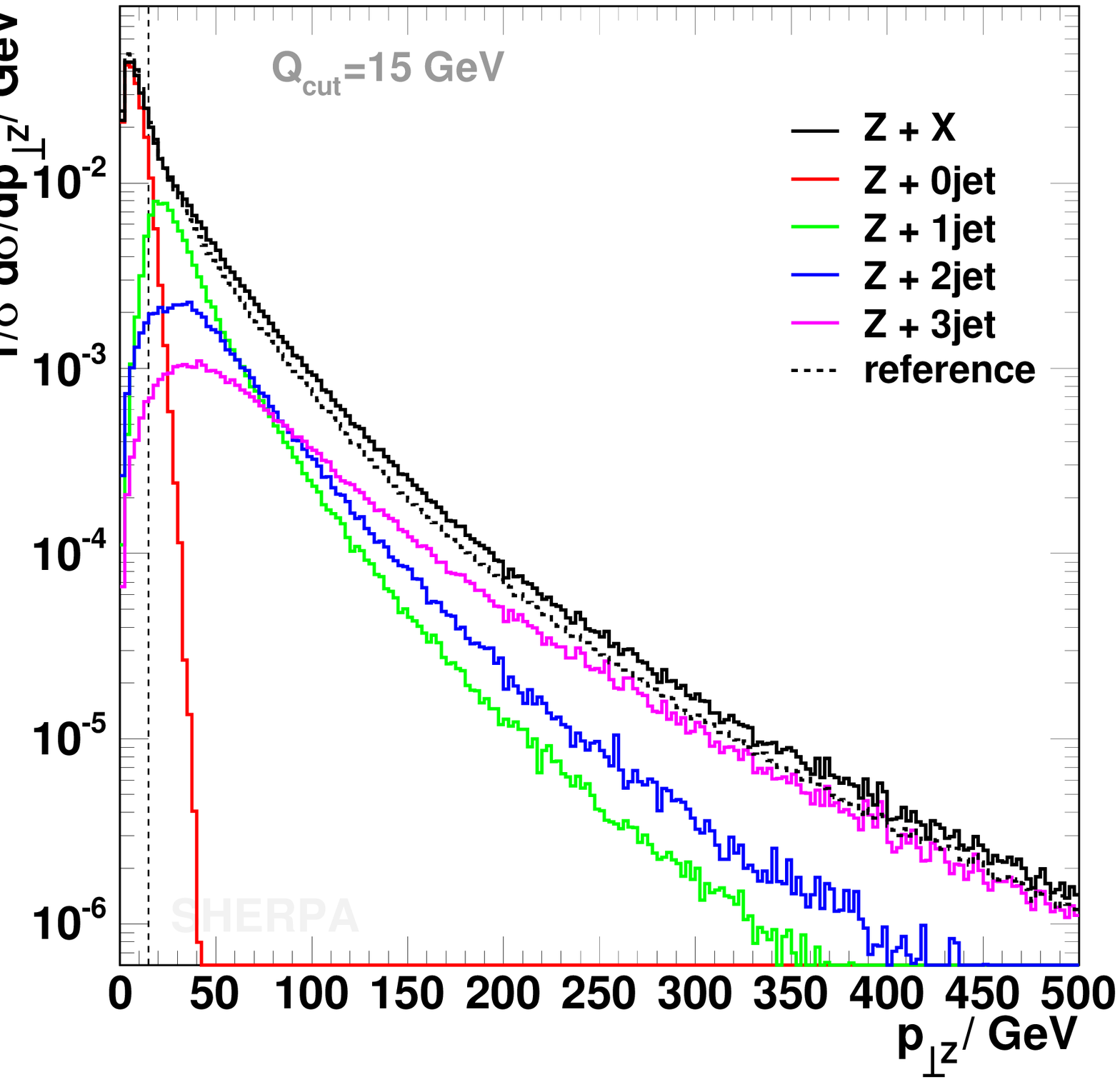}
\includegraphics[width=5.3cm]{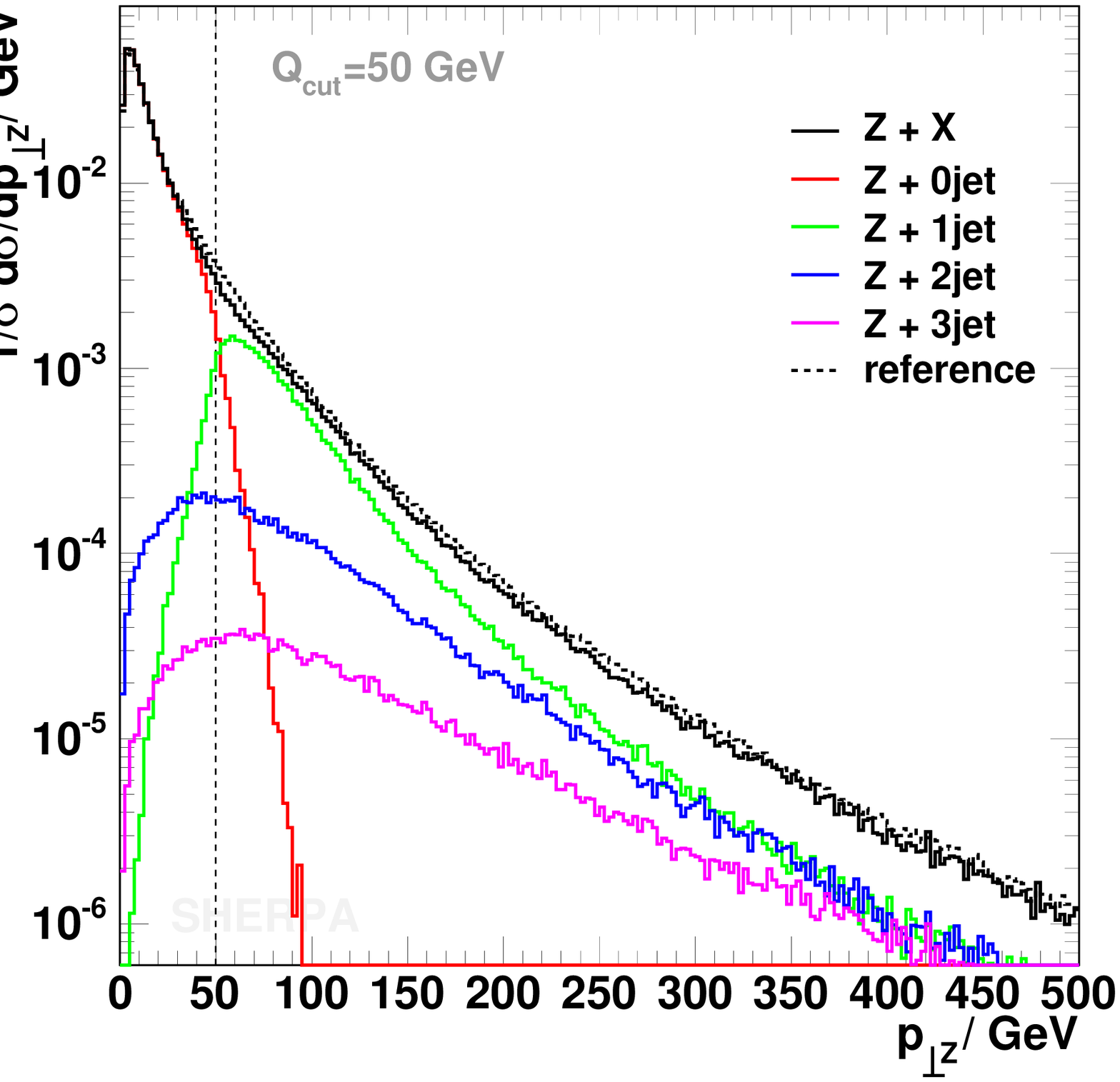}
\includegraphics[width=5.3cm]{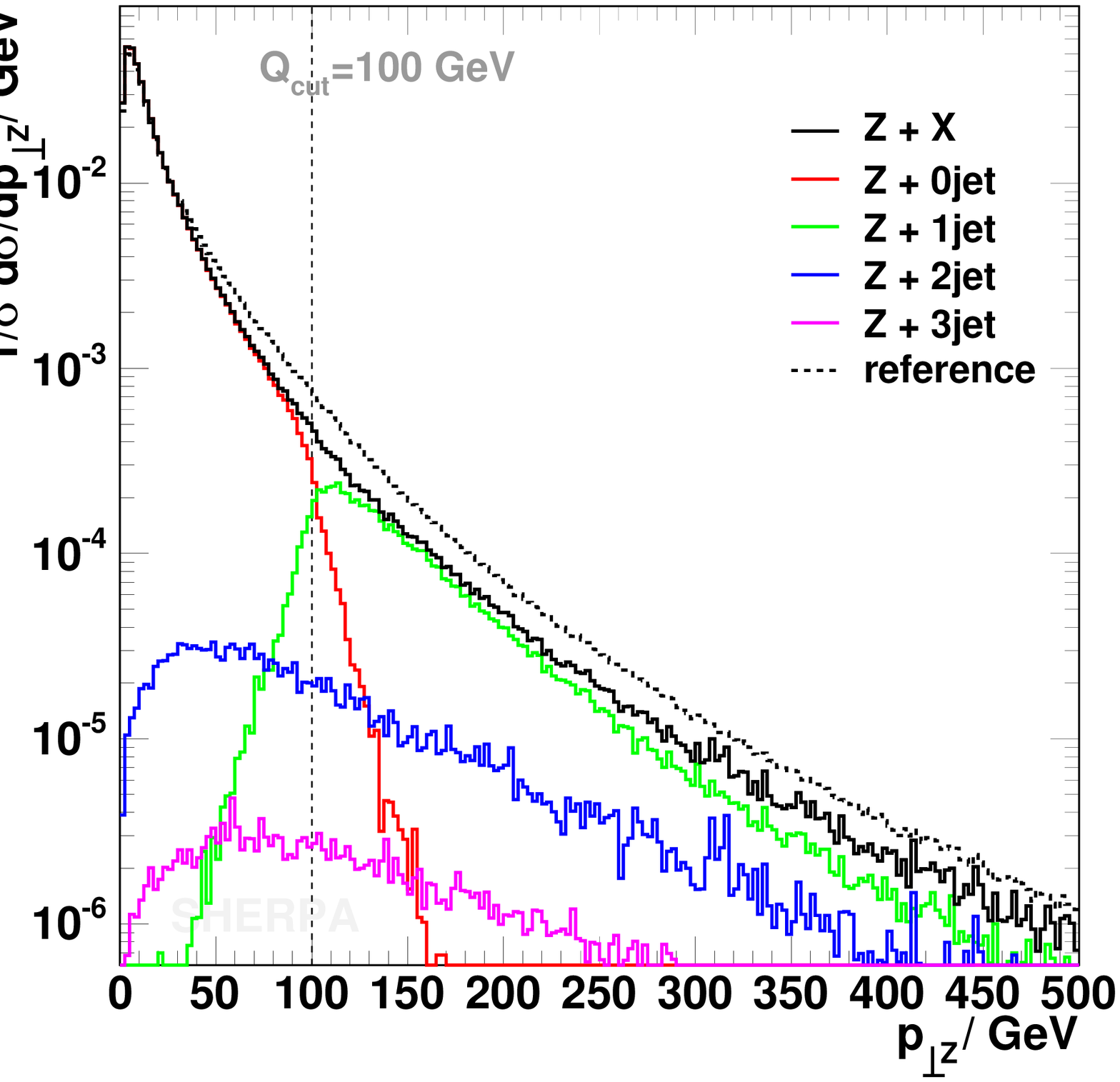}
\caption{\label{wjt:sherpa}
The $Z$ transverse momentum distribution, taken
from~\cite{Krauss:2005nu}.}
\end{figure}

Whichever matching procedure is used, an artificial parameter must be
introduced that serves to discriminate between the regions of hard
matrix elements and of collinear approximations. Although predictions
should be largely independent of this parameter, if it is not chosen
appropriately then the hard matrix elements may not be used when
necessary and the inadequacy of the original (unmatched) parton shower
returns. As an example of the matching procedure applied to $Z+$~jet
production, we refer to the CKKW scheme as implemented in the
SHERPA~\cite{Gleisberg:2003xi} event generator.
In this case the parameter is the jet resolution cutoff, $Q_{\rm cut}$,
and a study of its impact upon the inclusive $Z$ $p_T$ spectrum
is shown in Figure~\ref{wjt:sherpa}.
The contributions from each set of hard matrix elements are shown
separately, with the upper curve obtained from their sum. This curve
is not only smooth, but relatively stable with respect to the
change from $Q_{\rm cut}=15$~GeV to $Q_{\rm cut}=50$~GeV. However,
for $Q_{\rm cut}=100$~GeV the distribution begins
to fall off more rapidly at large $p_T$, indicating that the hard
matrix elements are not being effectively utilized.

As well as such studies of individual generators, recently
much work has gone into comparisons of the showering algorithms
and matching procedures that are in general use.
The results show that all generators are broadly consistent, with
differences that can mostly be accounted for by
the usual variation of scales~\cite{Alwall:2007fs}. Most importantly,
the input parameters can be tuned with Tevatron data and
then the results extrapolated for use at the LHC.

On the side of fixed-order calculations, the pace of developments that
{\em directly} confront experimental data has been somewhat slower.
The NNLO corrections to inclusive rates and distributions for $W$ and
$Z$ production have been known already for five
years~\cite{Anastasiou:2003ds}.
This accuracy, leaving residual uncertainties of a few percent, is
highly desirable -- for example, in order to pin down PDFs or possibly
even to serve as a luminosity
monitor~\cite{Dittmar:1997md,Khoze:2000db,Giele:2001ms}.
The NLO calculation for a final state involving one additional jet was
performed a long time ago~\cite{Giele:1993dj}, with the corrections to
$W/Z+2$~jet observables known at the same order more
recently~\cite{Campbell:2002tg}. The prospect of a NNLO calculation
of $W/Z+1$~jet production has become more imminent in the last year,
thanks to a pioneering effort to complete the NNLO corrections to
the 3-jet rate in $e^+ e^-$ collisions~\cite{GehrmannDeRidder:2007hr}.
The two processes are related
by crossing and, although it is non-trivial to address the difficult
issue of doubly-infrared singularities in the hadron collider
environment, a clear pathway between the two exists. For now an
extension of this approach to tackle the $W/Z+2$~jet calculations
at NNLO appears very unlikely and further calculational advances are
required.
Beyond two jets, fixed order calculations are limited to leading order
at present, although this may soon be addressed as
one of the plethora of approaches aimed at performing
arbitrary multi-leg processes at 1-loop order matures\footnote{
Indeed, there has been progress towards a calculation of
$W/Z+3$~jets at NLO since this symposium~\cite{Berger:2008sz}}.
However, for now one must choose
between either higher orders in QCD or a parton shower.

\begin{figure}
\includegraphics[width=8.5cm,viewport=0 463 449 737,clip]{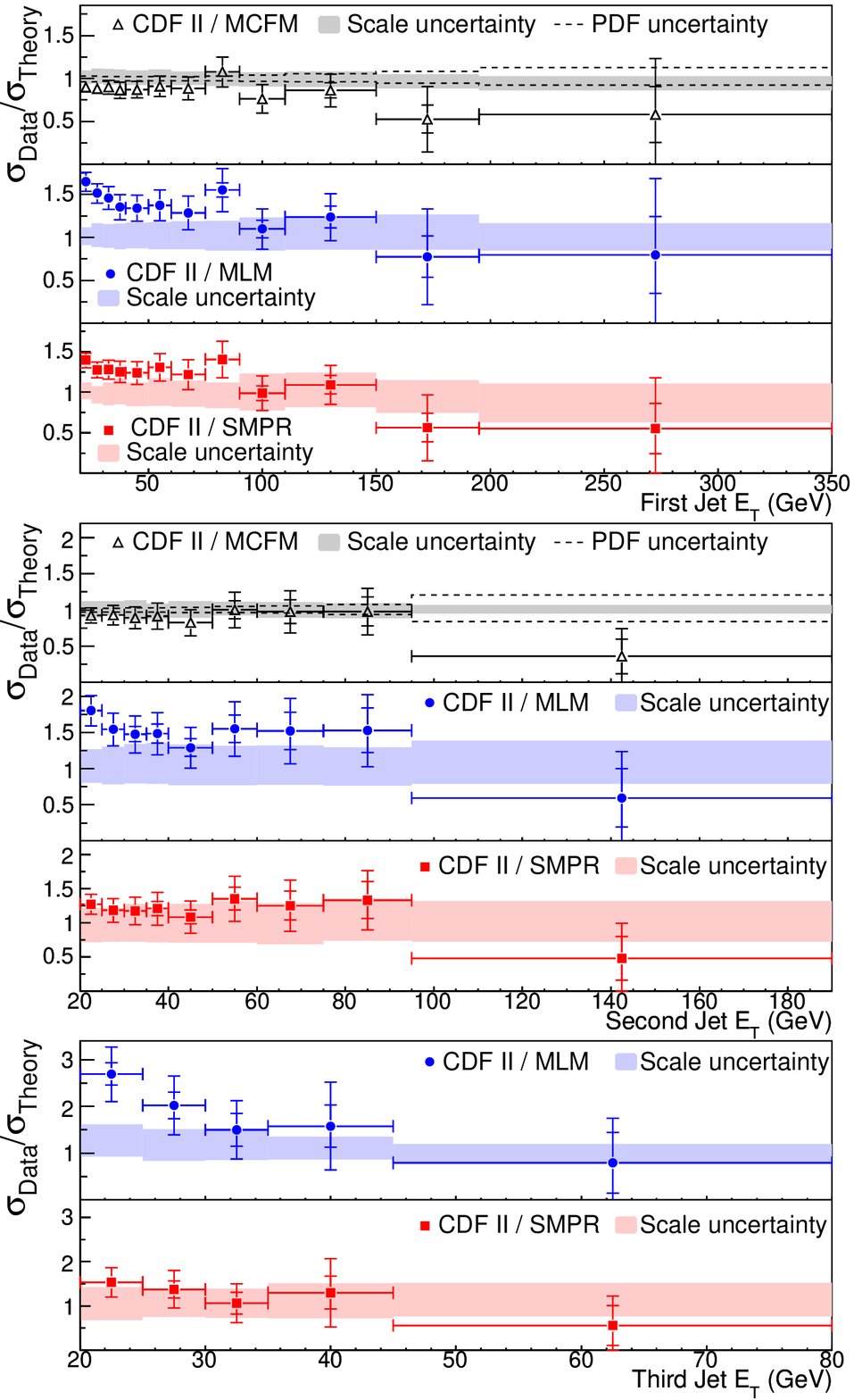}
\includegraphics[width=8.5cm,viewport=0 190 449 462,clip]{cdfwjets.eps}
\caption{\label{wjt:cdfwjets}
The $E_T$ of the first (left) and second (right) jets in $W+$jet events,
adapted from Ref.~\cite{Aaltonen:2007ip}. The CDF data are normalized to
NLO predictions (top), ALPGEN+HERWIG with the MLM matching prescription
(middle) and MADGRAPH+PYTHIA using the CKKW procedure (bottom).}
\end{figure}

A recent comparison of CDF $W+$~jet data with a variety of these
approaches is shown in Figure~\ref{wjt:cdfwjets}~\cite{Aaltonen:2007ip}.
It appears that, at least for
the first two jets, the parton shower approaches do not reproduce
the shape of the $E_T$ distributions very well. On the other hand,
the NLO calculation tracks the data reasonably successfully, with
deviations largely captured by the combination of scale and PDF
uncertainties. Of note is the omission of comparisons of the third and
higher jets; as already noted, for these no NLO prediction is currently
available and we must be content with qualitative agreement with
the parton shower approaches.

In light of this comparison, several open questions remain. The NLO
description appears to excel, but the agreement could be considered
too good. At the lowest values of $E_T$ considered one might
expect a fixed order calculation to be less accurate, but in fact this
is precisely where the NLO calculation agrees best and the parton
showers fare worst. At this point it is instructive to recall that
the NLO calculation cannot account for the effects of hadronization or
the underlying event. The lack of such information means that the
quality of the comparison may be somewhat misleading. This question
cannot be satisfactorily answered until a parton shower is developed to
NLO accuracy for this final state. We can hope that, by the time that
LHC analyses have reached a similar stage, such a tool is available.
A more immediate diagnostic of the fixed order and parton shower
approaches could be provided by a detailed comparison of predictions
for observables other than the jet transverse momenta. These may
indicate areas where the theoretical predictions perform particularly
badly and could be used to assess the degree to which the NLO agreement
might be accidental.

At the LHC similar studies will no doubt be performed. It is interesting
to note that related processes, such as $WW$ production, despite being
suppressed by a factor of 1,000 compared to the single $W$ process,
still have cross sections in the region of $100$~pb. The rates for
emitting additional radiation will again be suppressed by factors of
order $0.1$ for each additional jet. When combined with the large
luminosities expected at the LHC, this means that plenty of $WW+$~jet
events will be produced. Such events form an irreducible background
to various searches for the  Higgs boson when its mass is large enough
for the decay $H \to W W^*$ to be significant, notably in the gluon
fusion and vector boson fusion channels~\cite{Mellado:2007fb}.
In the former a jet veto is often applied to
reject other backgrounds, thus focussing on the $WW+0$~jets final state.
In the latter one expects the Higgs boson to be produced centrally,
accompanied by two very forward jets. Either both or just one of these
jets may be tagged (to increase the expected signal cross section),
thus requiring a good knowledge of $WW+1, 2$~jet observables. One could
therefore imagine that the breadth of experience accumulated in $W+$~jet
studies could profitably be applied to provide a systematic analysis
of $WW+$~jet backgrounds. On the
theoretical side, an extra vector boson means that parton shower (plus
matching) approaches will require more computational effort, but no
new developments. On the side of fixed order, it is quite a different
story. NLO corrections to inclusive $WW$ production have long been
known~\cite{Ohnemus:1991kk,Campbell:1999ah,Dixon:1999di},
with corrections to the emission of a single hard jet only very recently
computed~\cite{Dittmaier:2007th,Campbell:2007ev,Bern:2008ef}.
Again, the promised advances in multi-leg NLO computations could have
an immediate impact in this arena.

\section{$\mathbf W$/$\mathbf Z$ + HEAVY FLAVOR}

Although the production of heavy flavors (in particular, bottom
and charm quarks) and the production of
light quark jets share many similarities, at a theoretical level they
differ in a crucial way. For massless quarks, the matrix elements
diverge in the soft and collinear limits. This means that predictions
diverge, for example, in the limit that $p_T(q) \to 0$ so that
a minimum jet $p_T$ and angular separation must be imposed in order to
compare with similarly-restricted experimental data. For a massive
quark, its mass $M$ serves to regulate the divergence, which instead
behaves as $\log(M^2)$. Therefore no cuts are necessary in order to
properly define the cross section theoretically, so that it can then be
calculated inclusively. Unfortunately, this appealing feature comes at
a price: including the mass renders the fixed-order calculations more
difficult, which is particularly troublesome when already at the limit
of current sophistication -- namely, a vector boson plus two partons.

\begin{figure}
\includegraphics[height=12cm,angle=-90]{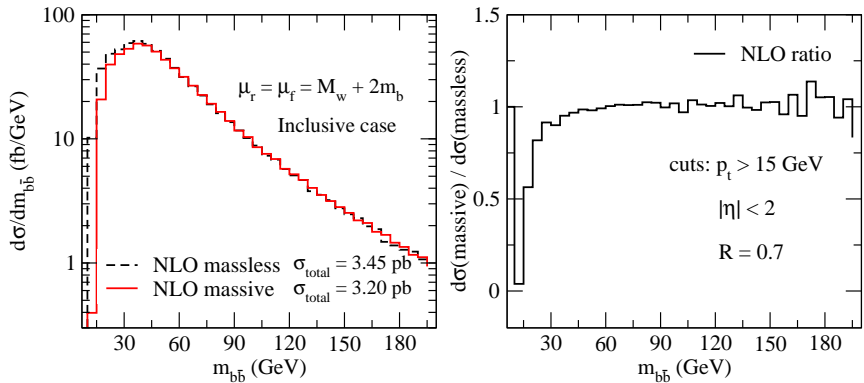}
\caption{\label{wjt:wbbcomp}
Left: A comparison of two NLO calculations of $Wb{\bar b}$ production
at the Tevatron, including the effects of the
$b$-quark mass (solid red) and neglecting them (dashed black).
The cross section is presented as a function
of the $b{\bar b}$ pair invariant mass, $m_{b\bar b}$.
Right: the ratio of the massive
and massless calculations, again as a function of $m_{b\bar b}$.}
\end{figure}

Of course, some analyses do not demand close attention to the low
transverse momentum behavior of these cross sections. In that case it
can be sufficient to work in the approximation that the heavy quark
is massless. The applicability of such an approximation is illustrated
well by the comparison between two calculations of $Wb{\bar b}$
production, as shown in
Figure~\ref{wjt:wbbcomp}~\cite{Febres Cordero:2006sj}.
When applying typical jet cuts, the most recent calculation that
includes the mass of the $b$-quark differs from an earlier one
for $m_b=0$ by less than $10\%$ in the total cross section
(i.e. a difference of order $m_b^2/p_T^{\rm min}(b)^2$).
As seen in the right panel, this deviation is
due entirely to the behavior in the threshold region that cannot
possibly be captured correctly in the massless calculation. Away from
threshold the two agree.

In some instances one cannot simply restrict the phase space in this
fashion. In particular, it is often advantageous to tag only one
heavy quark in order to improve experimental
efficiency (see for example Ref.~\cite{Acosta:2005ga}).
In such a scenario the partner anti-quark must be integrated over
the whole of the phase space region so that the straightforward
massless approach cannot be used. However another technique is
suitable in exactly this limit, one that is based on the use of
heavy quark parton distribution functions.

\begin{figure}
\includegraphics[width=3.7cm]{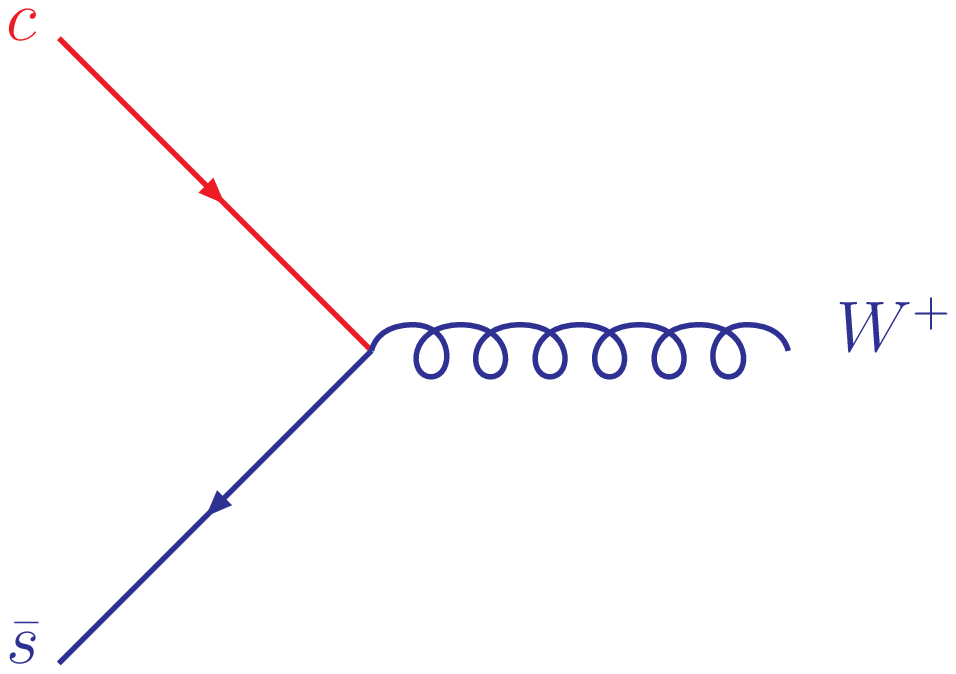} \hspace*{2cm}
\includegraphics[width=3.7cm]{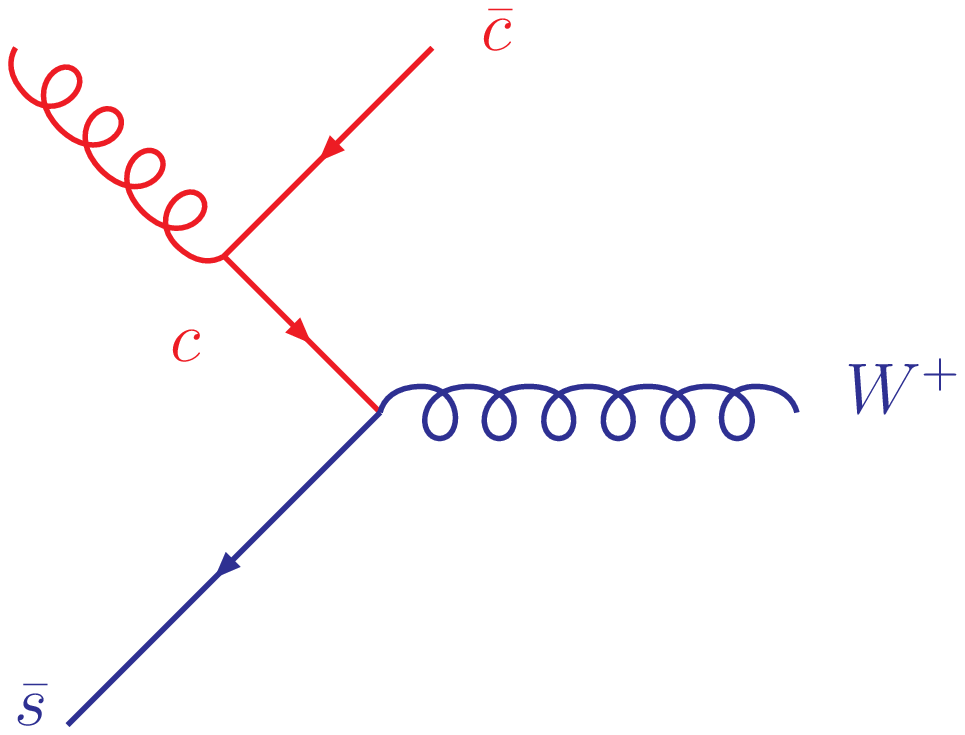}
\caption{\label{wjt:vfsffs}
Left: Production of a $W$ boson via the charm quark distribution
function, with an implicit anti-charm quark integrated out. Such a
diagram is included in the ``variable flavor scheme'' (VFS) when
the anti-charm quark is not relevant to the analysis.
Right: Production of a $W$ boson and an anti-charm quark via an
initial state gluon splitting, $g \to c\bar c$. If the anti-charm
quark is subsequently integrated out, this corresponds to the 
``fixed flavor scheme'' (FFS).}
\end{figure}

At very high center-of-mass energies we are used to the idea that we
may become sensitive to heavy quark content of the (anti-)proton sea.
In fact, we are well used to this approach in the context of $W$
production, where the contribution shown in the left diagram of
Figure~\ref{wjt:vfsffs}, dependent on the charm quark distribution,
is usually included when calculating the inclusive $W$ cross section
(and is responsible for about $5\%$ of the total at the LHC).
The bottom
quark distribution is a little less familiar, with its use at the
Tevatron mainly limited to studies of $t$-channel single top production.
However it will be more important at the LHC due to the increased role
of the proton sea. The heavy quark distribution function can be
thought of as folding back a gluon splitting into heavy quark pairs,
from the matrix elements and into the PDF. This splitting is shown
explicitly in the right-hand diagram of Figure~\ref{wjt:vfsffs}. The
calculation of these two processes, in the so-called variable flavor
scheme (VFS, left) and fixed flavor scheme (FFS, right), should
therefore agree at some level when information about the additional
massive anti-charm quark in the FFS is integrated out.

The two approaches are of course exactly equivalent in the full theory.
At a given order of perturbation theory this is not the case and,
in general, they do not agree particularly well at leading order.
Each approach has its own advantages and weaknesses, which are
summarized in Table~\ref{wjt:vfsffscomp}.
\begin{table}[h]
\begin{center}
\caption{A comparison of the features of calculations performed
in the variable and fixed flavor schemes, adapted
from~\cite{MManganoLBNL}.}
\begin{tabular}{|l|c|c|}
\hline Feature & Variable (heavy quark PDF) & Fixed (massive ME)
\\
\hline Recoiling heavy quark & no  & yes \\
\hline Exact kinematics      & no  & yes \\
\hline Higher order logs     & yes & no  \\
\hline Easier NLO            & yes & no  \\
\hline
\end{tabular}
\label{wjt:vfsffscomp}
\end{center}
\end{table}
In that light it is important to understand what differences exist
in practice and if (or when) one approach is superior. In particular,
parton showers typically make use of the fixed flavor scheme (for which
it is also easier to perform matching) so that the choice between the
two schemes is usually quite stark. 

\subsection{$\mathbf W$ + charm}

This is the simplest possible process for which the two schemes can be
compared~\cite{Berger:1988tu}. On the experimental side, the rate of events
is large and the final state is relatively easy to identify. On the
theoretical side, as previously noted, the VFS simply corresponds to
the $2 \to 1$ Drell Yan process (Figure~\ref{wjt:vfsffs}, left) that
is known to NNLO whilst the FFS calculation is a $2 \to 2$ scattering
(Figure~\ref{wjt:vfsffs}, right) that is known at
NLO~\cite{Giele:1995kr}.

Comparing the two calculations at LO, one finds that choosing the
``natural'' scale of the process, the $W$ mass (for both renormalization
and factorization scales), the cross section in the VFS is twice as
large as in the FFS~\cite{CMMTinprog}. However both calculations exhibit
strong scale dependence and, if one were to choose a small scale in
the region $(0.15 - 0.3) m_W$, the two would agree at
the $20\%$ level or better. Such a scale choice is well-motivated
theoretically, particularly for the VFS. One argument is that a more
appropriate scale for the gluon splitting that produces the charm
quark is the charm quark mass, so that a better single scale choice
is the geometric mean, $\sqrt{m_c m_W}$. Alternatively, one can argue
that the collinear approximation used in deriving the heavy
quark PDFs is only valid up to much smaller
scales~\cite{Plehn:2002vy,Maltoni:2007tc}. In any case, the VFS
result is largely independent of the choice of scale at NLO.

\begin{figure}
\includegraphics[width=5.5cm]{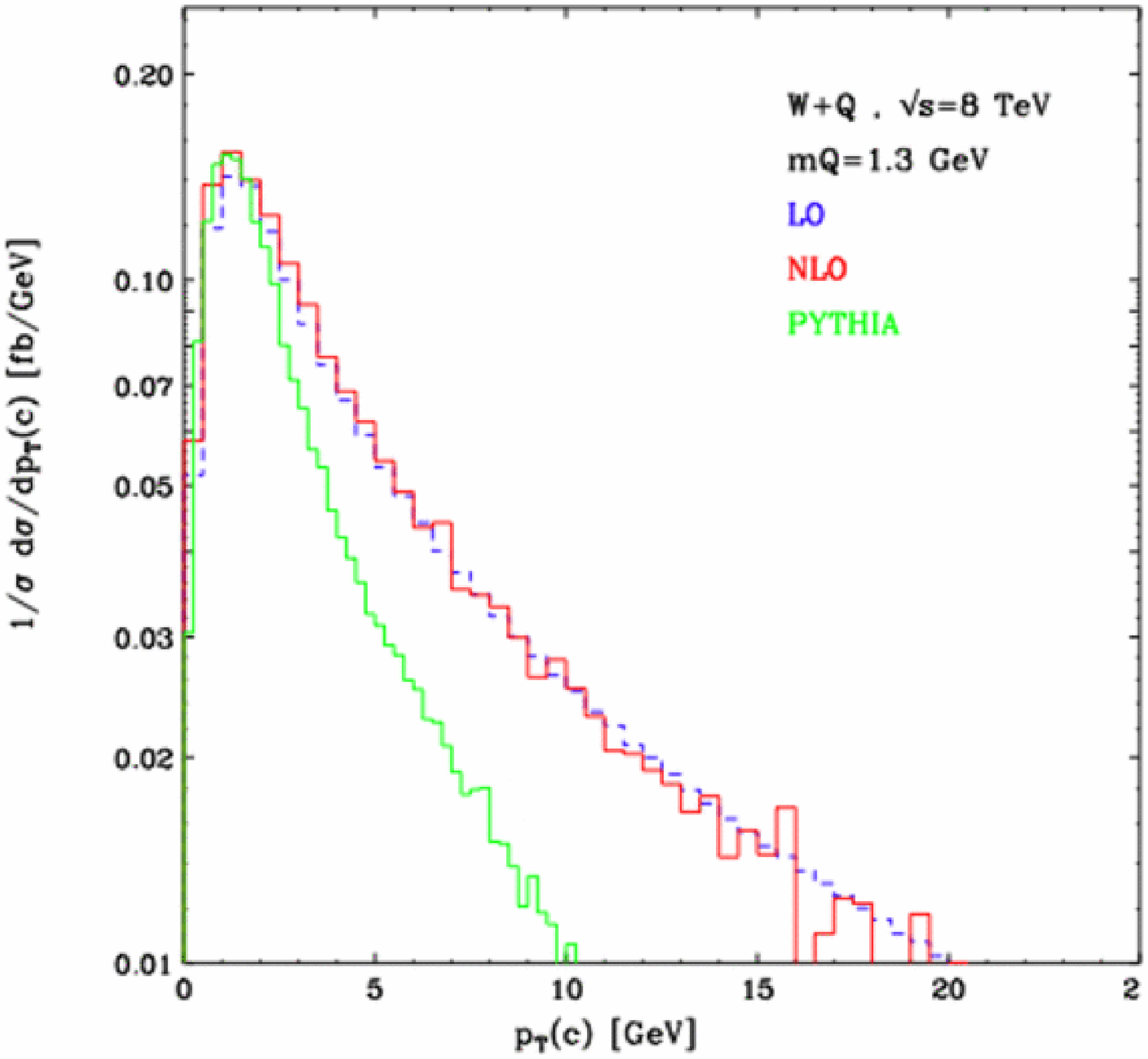}
\includegraphics[width=5.5cm]{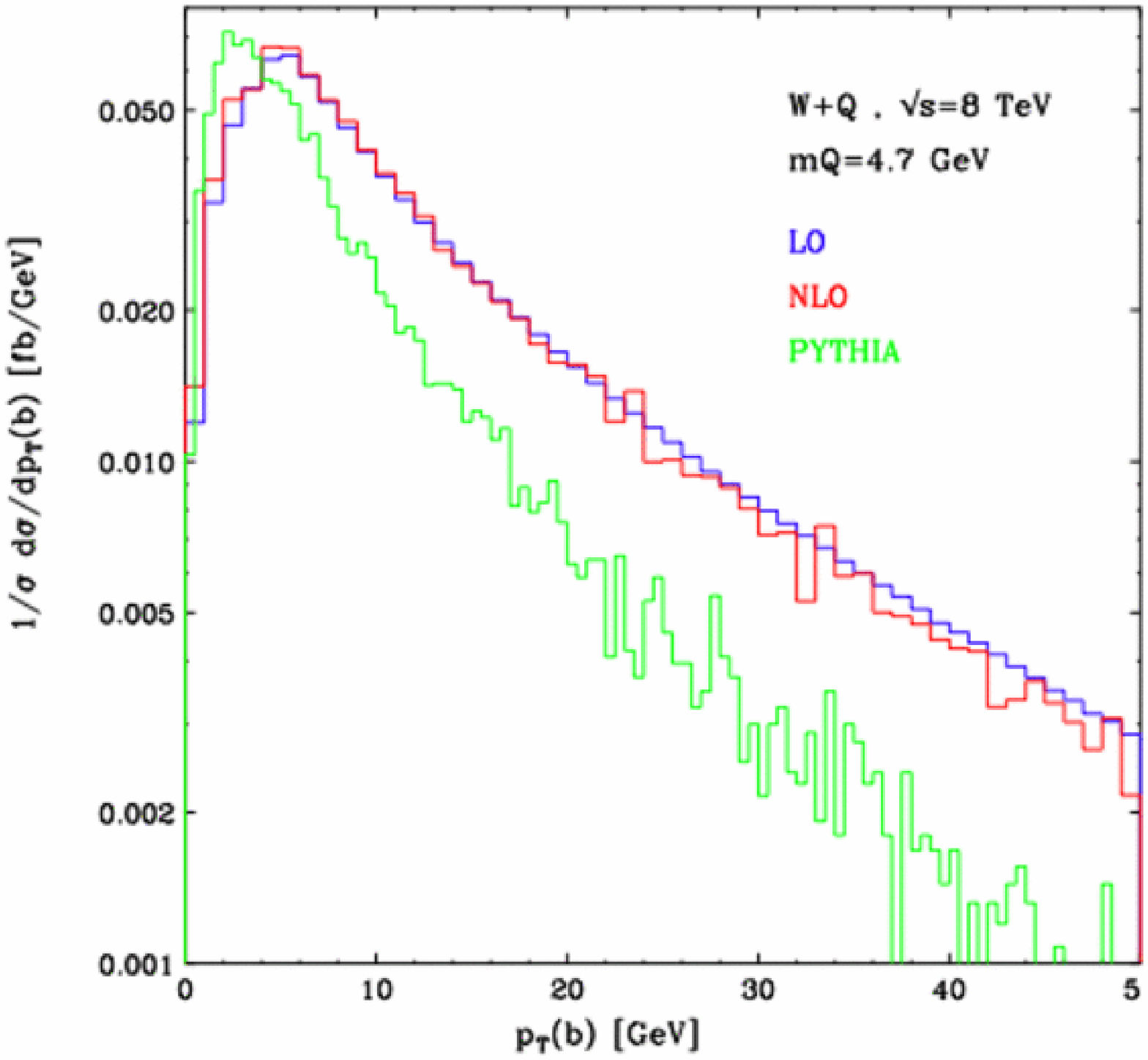}
\includegraphics[width=5.5cm]{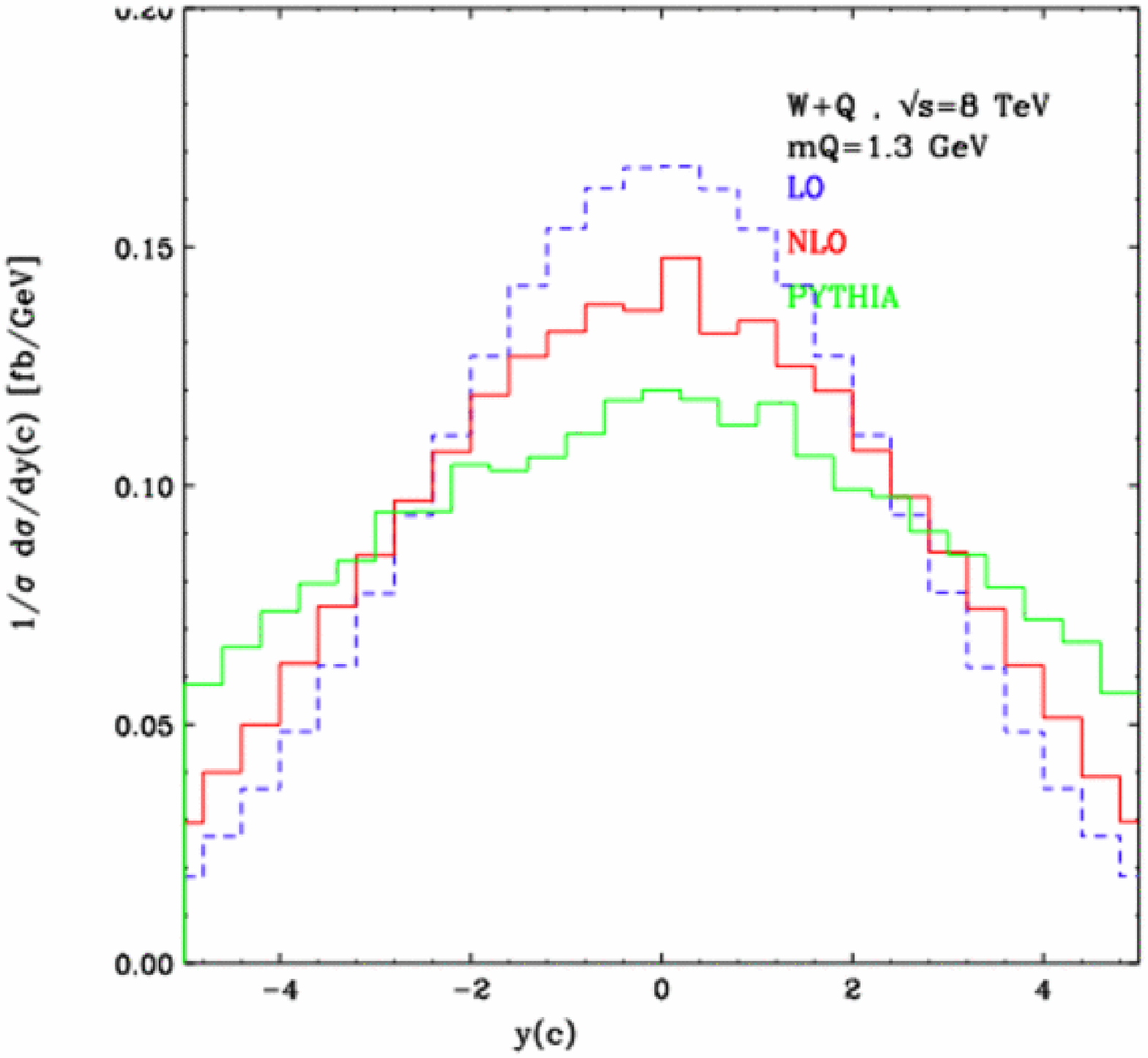}
\caption{\label{wjt:ffsvspythia}
Left: The $p_T$ of the charm quark predicted by the FFS (upper two
curves, overlapping) and in Pythia (lower curve).
Middle:The same quantity, but with the charm mass inflated to $4.7$~GeV
 in order to probe the kinematics involved in bottom quark production.
Right: The rapidity of the charm quark (FFS LO, upper; NLO, middle; 
Pythia, lower). All curves are normalized to the corresponding
predicted total cross section, so represent shapes only. 
Taken from~\cite{CMMTinprog}.}
\end{figure}

As previously noted, since the VFS
calculation yields approximately $5\%$ of the inclusive $W$ cross
section at the LHC, differences of this size between the two
calculations bring into question claims of a few percent accuracy of
the theoretical prediction. Even more important is the issue of
kinematical distributions, which are usually simulated using a parton
shower. A comparison between the predictions of the FFS
(at LO and NLO) with PYTHIA is shown in Figure~\ref{wjt:ffsvspythia},
for the transverse momentum and rapidity distributions of the charm
quark.
Firstly, it is clear from this figure that the prediction for the
shape of the transverse momentum distribution of the charm quark in the
FFS does not change from LO to NLO. The lack of substantial corrections
to this distribution at large $p_T(c)$ seems to indicate that there are
no large logarithmic corrections of the sort that purport to be resummed
in the VFS.  The PYTHIA prediction for the same
quantity of course lacks proper inclusion of hard radiation, leading
to a large underestimate at high $p_T(c)$. It is interesting
to note that the region of small transverse momentum appears to agree
with the FFS calculation (the left pane). However, when the charm
quark mass is increased in order to probe the kinematics involved in
bottom quark production (middle pane), the position of the peak at
low $p_T$ does not agree with the FFS curves. The underlying cause
of this disagreement is still under investigation, but the conclusions
could have significant implications for analyses at the LHC. Finally
it is heartening to note that, at least for the charm quark, the PYTHIA
prediction seems to reproduce one of the features of the NLO
calculation, namely a broadening in the rapidity distribution of the
charm quark (right pane). There is clearly much room for improvement,
particularly as regards to parton shower/matrix element merging, which
should shed further light on these comparisons.

Of course, the ultimate arbiter in such disputes is the data itself.
A recent result from CDF, with the very basic selection cuts
$p_T(c) > 20$~GeV, $|\eta(c)| < 1.5$, yields the result,
$\sigma \times \mathrm{BR}(W \to e\nu) = 9.8 \pm 3.2$~pb~\cite{
Aaltonen:2007dm}. This is to be compared with the NLO prediction
in the FFS of $\sigma(\mathrm{NLO}) = 11.0 (+1.4-3.0)$~pb, where the
central result corresponds to a scale of $Q=40$~GeV and the theoretical
error accounts for variation both over a wide range of scales and over
a selection of PDF sets. Thus there is good
agreement between experiment and theory, albeit with large residual
uncertainty on both sides.
 
\subsection{$\mathbf W$ + bottom}

\begin{figure}
\includegraphics[width=4cm]{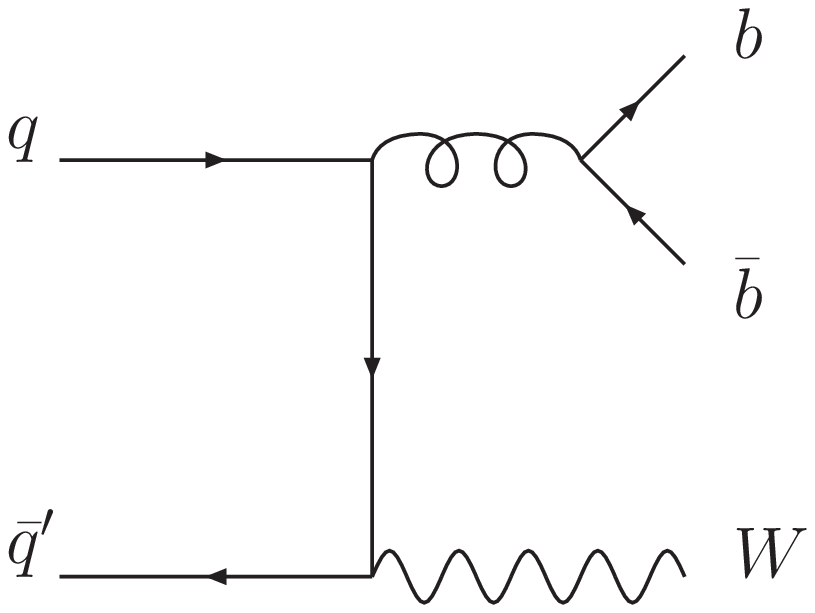}\hspace*{1cm}
\includegraphics[width=4cm]{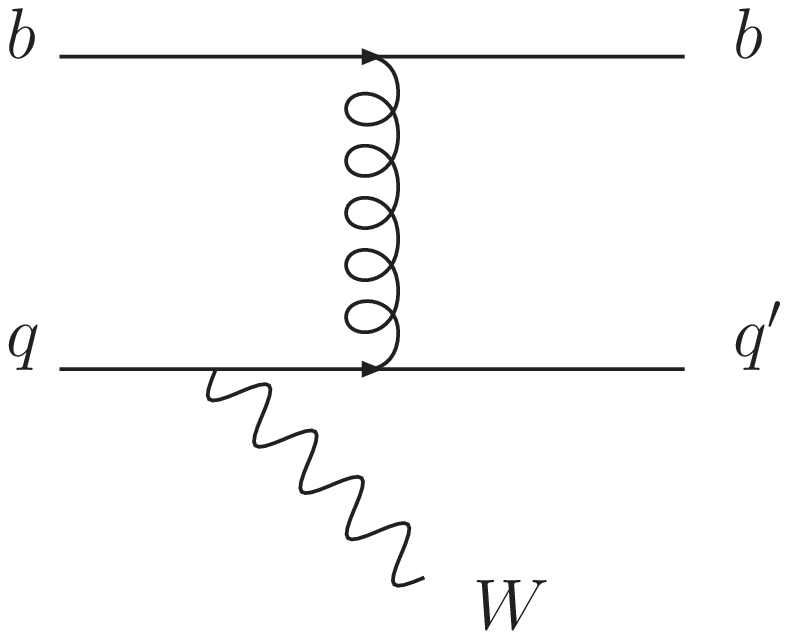}
\caption{\label{wjt:wbdiags}
Left: LO diagram for production of a $W$ and two $b$-quarks in the FFS.
Right: LO diagram for production of a $W$, a $b$ quark and a light
quark (VFS).}
\end{figure}

This channel is not a direct analogue of $W+c$ production, due to
CKM suppression. Instead, a hard bottom quark must be produced together
with other (undetected) particles. The two leading mechanisms for
producing such final states are shown in Figure~\ref{wjt:wbdiags}.
The left pane shows usual $Wb{\bar b}$ production (in the FFS), where
we assume that only one of the quarks is tagged. As discussed above,
this requires a calculation including the bottom quark mass, which is
known at NLO~\cite{Febres Cordero:2006sj}. The right pane shows 
production of the state $Wbq$ (with $q$ a light quark) in the VFS,
which is also known at NLO~\cite{Campbell:2006cu}. By demanding that the
$b$ quark is observed, no singularities are encountered when evaluating
this contribution. Note that the NLO corrections to the corresponding
FFS process, a $Wb{\bar b}q$ final state, are not known.

To obtain the best possible prediction for the rate of $Wb$ production,
both of the above contributions can be combined. Care must be
taken to ensure that no overlap between the calculations is included,
but some preliminary results are already available~\cite{CEMWFRW}. The
relative importance of the two contributions varies with the collider:
at the Tevatron the valence quarks ensure the $Wb{\bar b}$ process
dominates, whereas at the LHC the gluon (and hence $b$-quark)
flux means the $Wbq$ process is largest. Early results indicate that,
for typical cuts, the total cross section is enhanced by a factor of
about $1.5$ at both machines. Although this is a substantial increase
it is not enough to explain preliminary results from
CDF~\cite{CDFwb} that indicate a shortfall of about a
factor of three or four when comparing LO theory to data.
This is especially intriguing since
the shapes of distributions are reasonably well described. At this
stage it seems that the NLO corrections may be part of the
solution, but are unlikely to solve the puzzle completely.

\subsection{$\mathbf Z$ + heavy quark}

\begin{figure}
\includegraphics[width=4.5cm]{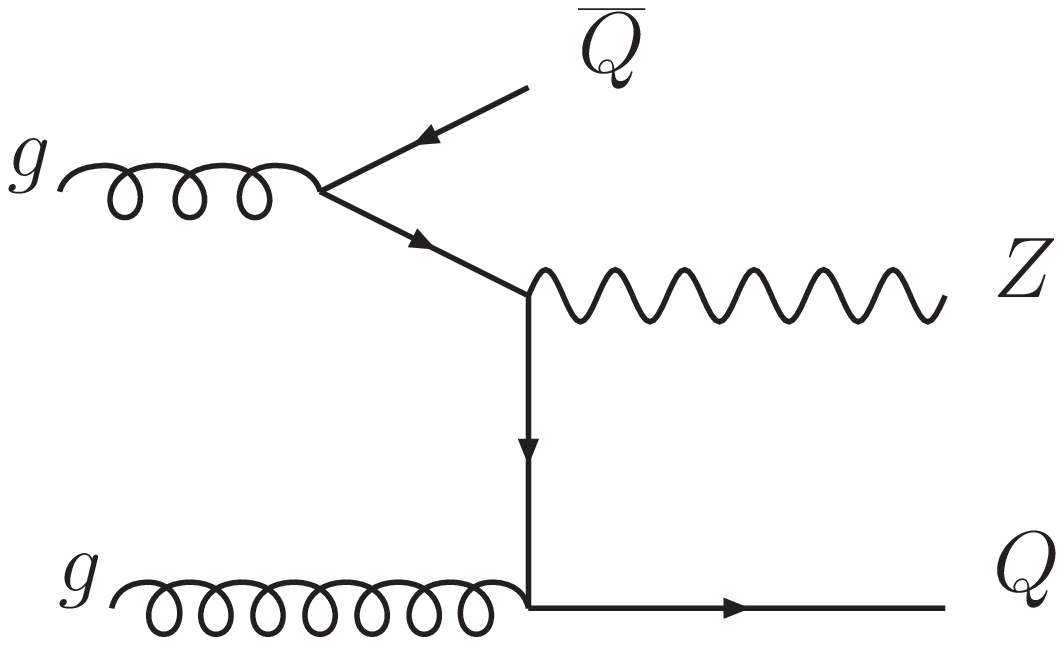}\hspace*{1cm}
\includegraphics[width=4.5cm]{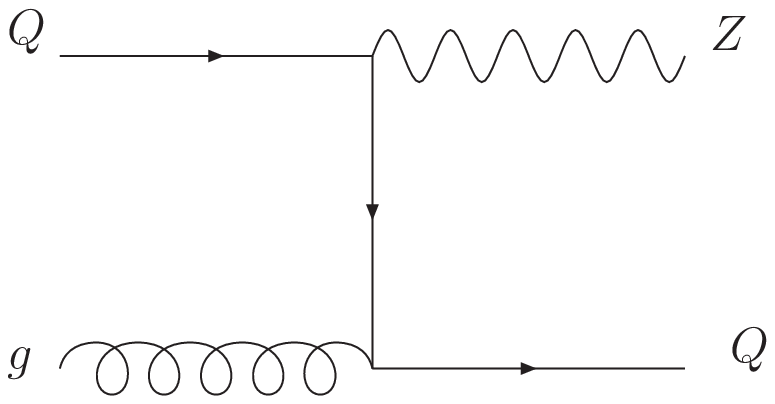}
\caption{\label{wjt:zhq}
Left: LO diagram for the process $gg \to ZQ{\bar Q}$ (FFS).
Right: The equivalent process using a heavy quark PDF (VFS).}
\end{figure}

\begin{figure}
\includegraphics[width=4.5cm,angle=90]{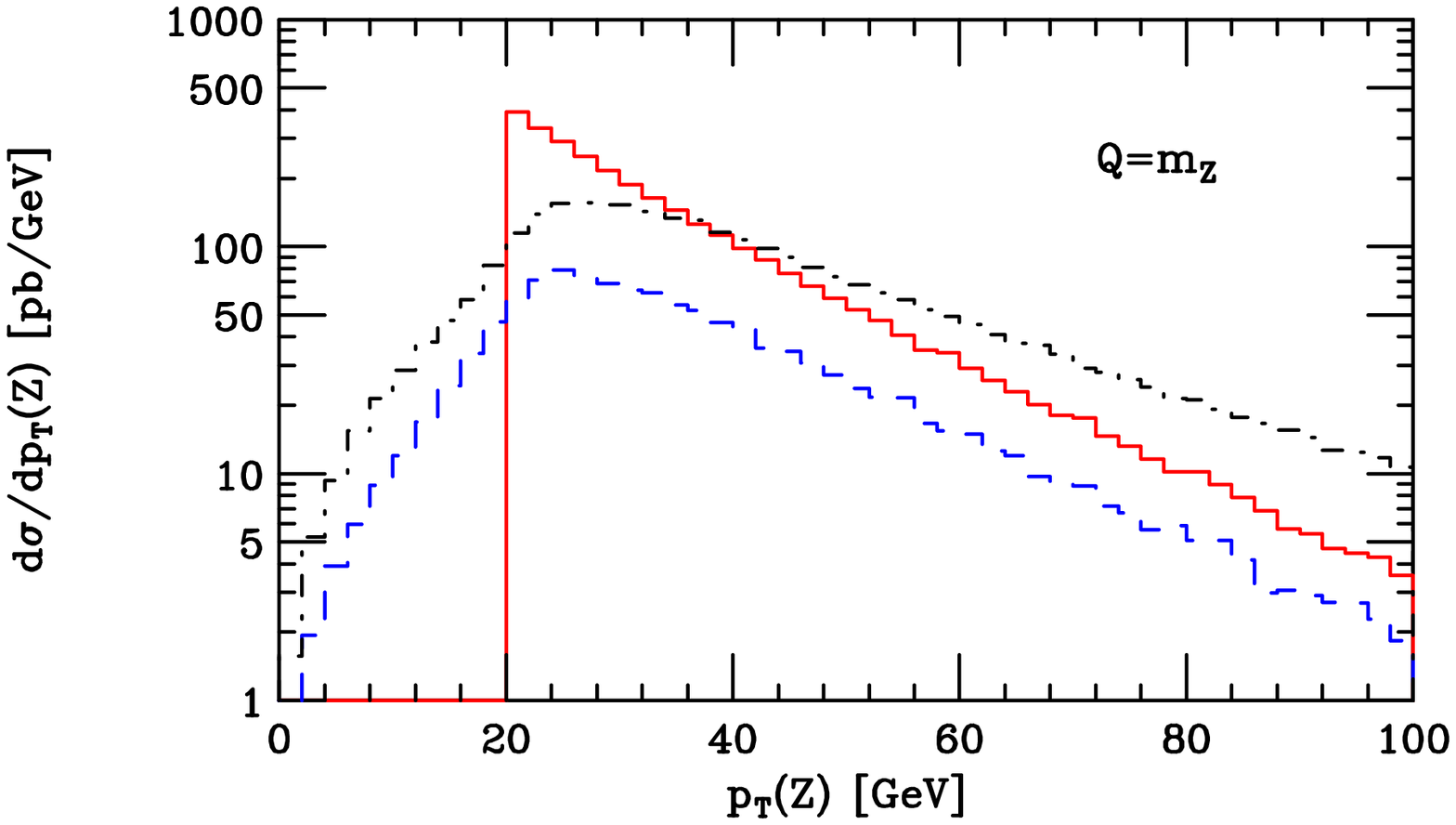}
\includegraphics[width=4.5cm,angle=90]{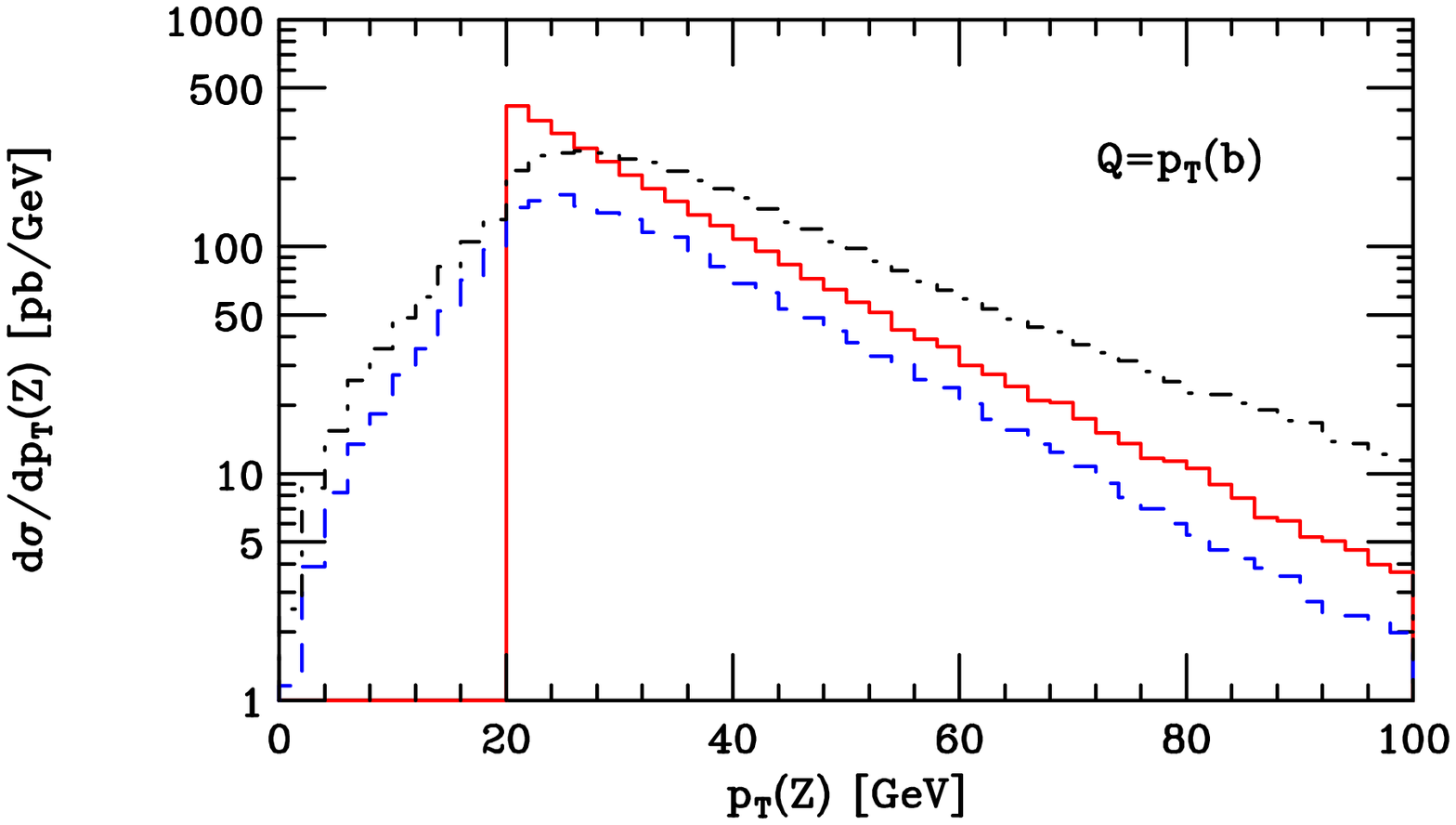}
\caption{\label{wjt:zhqptz}
The transverse momentum distribution of the $Z$ boson produced in
$Z+b$ final states as calculated in the FFS and VFS using a high
scale $Q=m_Z$ (left) and a much smaller one, $Q=p_T(b)$ (right).
The VFS (red, solid) curve is zero in the region below the kinematic
limit, $p_T(Z) < 20$~GeV. The lower (blue, dashed) curve is the FFS
result. The upper (black, dot-dashed) curve shows the contribution
from the other $q{\bar q} \to Zb{\bar b}$ process that is common
to both schemes.}
\end{figure}

The production of a $Z$ boson in association with either a bottom or
charm quark is somewhat analogous to $Wb$ production. The additional
complication is that the FFS now also receives a contribution from
$gg$ initial states at LO, which can alternatively be described by a
simpler VFS calculation (see Fig.~\ref{wjt:zhq}).
A comparison of the two approaches (at LO) is shown in
Fig.~\ref{wjt:zhqptz}, using the cuts $p_T(b)>20$~GeV, $|\eta(b)|<2.5$
at the Tevatron.
There are substantial differences between the two calculations,
which are somewhat reduced for a smaller scale choice.
Now that the NLO corrections to both processes are
known~\cite{Campbell:2003dd,FebresCordero:2008ci}, real understanding
may follow from a rigorous comparison of the two.
The observed discrepancy is in marked contrast to the (similar) process
in which a photon and bottom quark are observed in the final state. In
that case the scale of the problem is simply set by the photon $p_T$
and a comparison of the two schemes at LO already shows good
agreement~\cite{MManganoLBNL}.

Once again, experimental data has the chance to discriminate between
the two calculational schemes. However, information on the total
rate obtained by the D0 experiment~\cite{Abazov:2004zd}
does not immediately yield the desired information. The ratio
${\cal R}= \sigma(Z+b-\mathrm{jet})/\sigma(Z+\mathrm{jet})$
is measured to be ${\cal R}_\mathrm{exp} = 0.021 \pm 0.004$,
to be compared with the predictions,
${\cal R}_\mathrm{VFS, NLO} = 0.018 \pm 0.004$ and
${\cal R}_\mathrm{FFS, LO} = 0.015$--$0.023$.
The reason for the good agreement between the two calculations can be
found in Fig.~\ref{wjt:zhqptz}. It is clear that the
$q{\bar q}$ contribution, present in both schemes, dominates the total
cross section. It is only by comparing the $p_T$ distribution of the
$Z$ below about $40$~GeV that the two schemes could be probed. It is
essential that such analyses are performed in order to better understand
the theoretical predictions.

\section{OUTLOOK} 

In the absence of heavy quarks, the situation is quite encouraging.
There is good agreement between the latest experimental results and
theoretical predictions. The technique of matrix element matching in
a parton shower is by now a mature field and, if anything, could
benefit from more experimental measurements with which to compare and
tune. With regards to NLO, results for $W$ and $Z$ production with up to
two hard jets agree well with data, but cannot address
higher-multiplicity final states. However a host of new automated
multi-leg approaches should fill this need in the near future.

When considering heavy quark production, the picture is not so clear.
In some cases (notably $Wb$ production) theoretical predictions fail
to reproduce the rates observed experimentally; elsewhere agreement is
patchy at best. However it is heartening to see that data from the
Tevatron is now seriously confronting the two theoretical approaches
commonly used in this arena, the FFS and VFS. On the theoretical
side there has been a wealth of activity in recent years, summarized
in Table~\ref{wjt:activity}.
\begin{table}[t]
\begin{center}
\caption{\label{wjt:activity}
Theoretical activity in $W/Z+$~heavy flavor in recent years.}
\begin{tabular}{|l|c|c|c|c|}
\hline & 1 $c$-tag & 1 $b$-tag & 2 $c$-tags & 2 $b$-tags
\\
\hline $W+1$~jet  & FFS NLO~\cite{Giele:1995kr,Campbell:2005bb} & FFS+HVQ NLO~\cite{CEMWFRW} & n/a & n/a \\
\hline $W+2$~jets & LO only & HVQ NLO~\cite{Campbell:2006cu} & \multicolumn{2}{c|}{FFS NLO~\cite{Febres Cordero:2006sj}} \\
\hline $Z+1$~jet  & \multicolumn{2}{c|}{FFS NLO~\cite{FebresCordero:2008ci}, HVQ NLO~\cite{Campbell:2003dd}} & n/a & n/a \\
\hline $Z+2$~jets & \multicolumn{2}{c|}{HVQ NLO~\cite{Campbell:2005zv}} & \multicolumn{2}{c|}{FFS NLO~\cite{FebresCordero:2008ci}} \\
\hline
\end{tabular}
\end{center}
\end{table}
This opens up a real chance to make a
systematic evaluation of these calculations, both with each other and
against data, in order to properly interpret the theoretical tools.

\end{document}